\documentstyle[aps]{revtex}
\input{psfig}
\title{Lorentz Covariant Spin--Grouping of Baryon Resonances}
\author{M.\ Kirchbach
\thanks{ {\it E-mail address}: mariana@kph.uni-mainz.de}}
\address{Institute for Nuclear Physics, University Mainz,
D-55099 Mainz, Germany}
 
\begin{document}

\maketitle

\begin{abstract}
A well pronounced spin--grouping of baryon resonances 
to O(4) partial waves is
found in baryon spectra and shown to be well interpreted
in terms of Lorentz group representations 
of the type $\lbrace {1\over 2}+l', {1\over 2}+l'\rbrace
\otimes \lbrack\lbrace {1\over 2},0 \rbrace \oplus
\lbrace 0,{1\over 2}\rbrace \rbrack $
with $l'$ integer. In this way the relativistic description
of finite dimensional resonance towers containing
higher-spin states becomes possible. It is further argued that
the nucleon excitations into the $l'=1$ and $l'=2$ multiplets
are chiral phase transitions.
\end{abstract}

\vspace*{0.25truein}

\noindent
The relativistic description of resonances with higher-spins is
one of the oldest unsolved problems of baryon spectroscopy.
Attempts to handle it in the mid to late 60's failed through 
the improper choice for the irreducible representations of the Lorentz 
group considered to embed the baryon excitations known at that time.
For example, Kleinert considered in \cite{Kleinert} unitary 
representations of the Lorentz group describing a spin-tower of 
states without upper end starting for the nucleon $P_{11}(939)$ with
the $D_{13}(1520)$, $F_{15}(1680)$, $G_{17}(2190)$ etc. resonances.
Alternatively, Regge phenomenology required in the t-channel near 
origin (i.e. t$_\mu \approx $ 0) a grouping of approximately
mass degenerate baryon Regge poles
into the finite dimensional non-unitary representations of the 
Lorentz group of the type $\lbrace l_1, l_2\pm {1\over 2} \rbrace $ with 
integer $l_i $'s (so called Toller poles \cite{Toller}). Unfortunately, 
both the schemes mentioned above are in disagreement with data, in particular 
because the parity duplication of the resonances required by them is absent 
from the spectra. On the other side, Weinberg described baryon resonances as 
isolated $\lbrace J,0\rbrace \oplus \lbrace 0,J\rbrace $ higher--spin states. 
In order to obtain the corresponding relativistic equations of motion Weinberg 
put forward the idea in \cite{J00J} to embed the 
$\lbrace J,0\rbrace \oplus \lbrace 0,J\rbrace $ state,
considered as pointlike, within the Rarita--Schwinger multiplet 
$\lbrace {1\over 2}\, k,{1\over 2}\, k\rbrace \otimes $
$\lbrack \lbrace {1\over 2},0\rbrace $
$\oplus \lbrace 0,{1\over 2}\rbrace\rbrack $ (with $k$ integer and related to $J$ 
via $k=J-1/2 $ ) and to eliminate all the lower--spin components, considered as 
{\it redundant unphysical states\/} by means of a set of suitably choosen 
auxiliary conditions. Within Weinberg's scheme a spin--J resonance of mass $M$ and 
momentum $p_\mu $ is described by a totally symmetric traceless rank--$k$ 
Lorentz tensor with Dirac spinor components
$\Psi_{\mu\mu_1 ...\mu_{k-1}}$ satisfying the conditions
\begin{eqnarray}
(\gamma\cdot p -M)\, \Psi_{\mu \mu_1...\mu_{k-1}} &=&0\, ,\\
p^\mu \Psi_{\mu \mu_1...\mu_{k-1} } =0\, ,
&\quad &
\gamma^\mu \Psi_{\mu \mu_1...\mu_{k-1}} =0\, .
\label{Wbrg_spinor}
\end{eqnarray}
Here, Proca's equation in Eq.~(\ref{Wbrg_spinor}) 
eliminates the spin--0 and preserves
the spin--1 components from the $\lbrace 1/2,1/2\rbrace $ representations 
as associated with the Lorentz indices. The second set of auxiliary conditions 
in (\ref{Wbrg_spinor}) eliminates then the longitudinal components of the spin--1 
field. It ensures that the state surviving the elimination procedure carries the 
maximal allowed helicities and describes, therefore, the highest--spin 
state of the Rarita--Schwinger multiplet. 
A serious disadvantage of the Rarita--Schwinger scheme for treating 
higher--spin states is that for 
off--shell particles the auxiliary conditions are not any longer valid
and the appearance of arbitrary unphysical parameters in the 
Lagrangian describing the couplings of the states under
consideration to external fields is inevitable.
Because of all these difficulties, the requirement 
for relativistic and thereby for field theoretical description of higher--spin 
resonances was gradually given up by hadron spectroscopy thus opening space for 
pragmatic evidently non--relativistic schemes such as the 
O(3)$_L\otimes $SU(6)$_{SF}$ hadron book--keeping symmetry.
Through this group the trivial spin-flavor $(SF)$ correlation
between three quarks in the 1s-shell was naively extended to arbitrary 
orbital angular momenta. The problems raised by symmetry groups of strong 
interaction based on tight correlations between the spin- and flavor 
degrees of freedom of quarks not only have 
conceptual disadvantages, but also their predictions are not
satisfactory. For example, the O(3)$_L \otimes$SU(6)$_{SF}$
classification scheme predicts a substantial excess of
baryon excitations known as `missing resonances'.
Finally, the basic idea of the multiplets 
as well separated families of particles of 
different internal but same space--time properties, 
seems quite inappropriate here, where 
the spacing between the O(3)$_L \otimes$SU(6)$_{SF}$ multiplets is
much smaller as compared to the mass splitting within the multiplets.

On the other side, speed plot analysis of the pole positions 
on the complex energy plane of various baryon resonances  ($L_{2I, 2J}$) 
with masses below $\sim $ 2500 MeV performed by H\"ohler~ \cite{Hoehler} 
revealed a well-pronounced spin--clustering in baryon spectra.
This is quite a surprising result as it was not anticipated by any model or 
theory. Independently, the symmetry of all reported $N$, $\Delta $ and 
$\Lambda $ baryon excitations with masses below 2500 MeV was re-analyzed in 
our previous work \cite{Ki1} and shown to be governed by 
O(1,3)$_{LS}\otimes $SU(2)$_I$ rather than by 
O(3)$_{L}\otimes $ SU(6)$_{SF}$ as predominantly used 
since the invention of the naive three flavor quark model.
The O(1,3)$_{LS}\otimes $SU(2)$_I$ symmetry indicates that
the spin-orbital correlation between quarks is much stronger than the 
spin-flavor one. Indeed, it was demonstrated in \cite{Ki1}
that H\"ohler's poles correspond to O(4) partial waves
as they are identical to Lorentz multiplets of the
type $\lbrace {1\over 2}+l' ,{1\over 2}+l' \rbrace \otimes 
\lbrack \lbrace 1/2,0\rbrace \oplus \lbrace 0, 1/2\rbrace\rbrack $
with $l' $ integer. All orbital angular momenta (denoted here by $l$) contained 
within a $\lbrace {1\over 2}+l' ,{1\over 2}+l' \rbrace $ multiplet
have either natural or unnatural parity
\footnote{The parity $(-1)^{L+1}$ of 
a single $\pi N$ resonance $L_{2I,2J}$ in standard notation \cite{Part},
where $L$ stands for the relative angular momentum in the
decay channel, is determined in the present classification scheme by either 
$(-1)^l$ or $(-1)^{l+1}$, depending on whether the parity of the intrinsic orbital 
angular momentum is natural or unnatural. In the present notation, 
L takes the values of either $L=|l-1|, (l+1)$ for natural, or $L=l$ for unnatural 
parities.}. Coupling a Dirac particle to $l$ is then standard and leads to covariantly 
transforming  spin--groups of states with  
$\vec{J}=\vec{l} \otimes \vec {1\over 2}$ and $l=0, ...,2l'+1$ 
all having either natural or unnatural parities according to 
$J  = {1\over 2}^+,{1\over 2}^-,{3\over 2}^-,...,
(k+{1\over 2})^\pi $ with $k=2l'+1$, $\pi = (-1)^l$  for natural, and 
$ \pi = (-1)^{l+1}$ for unnatural parities.
Note that the $\lbrace {1\over 2}+l', {1\over 2} +l'\rbrace $ multiplets
are well known from the Coulomb problem, where 
they correspond to even principal quantum numbers $n=  k+1$.
These finite dimensional unitary representations of O(4), the maximal
compact group of the Lorentz group, 
contain only once each O(3) representation corresponding to
an intrinsic orbital angular momentum $l=0, ..., k$. 
The Coulomb multiplets  yield in the O(1,3) group
finite dimensional {\it non--unitary} representations.
Because of the possibility of restricting O(1,3) to
$SU(2)\otimes SU(2)$ for the $\lbrace {1\over 2}+l' ,{1\over 2}+l' \rbrace $
representations, however, the latter may be used for the construction
of relativistic equations of motion.
The baryon resonances are now collected into the following
representations \cite{Ki1}: 
\begin{eqnarray}
l'=0:\quad \Psi_\mu \, &:& P_{2I,1}; S_{2I,1}, D_{2I, 3}\, ,
\quad \mbox{for}\quad I= 0,\, {1\over 2},\, {3\over 2}\, ,
\quad \mbox{and}
\nonumber\\
l'=1:\quad \Psi_{\mu\mu_1\mu_2}\, &:&
S_{2I,1};P_{2I,1}P_{2I,3};D_{2I,3},D_{2I,5}; F_{2I,5}, F_{2I,7}\, ,
\nonumber\\
l'=2:\quad \Psi_{\mu\mu_1...\mu_4} \, &:&
S_{2I,1};P_{2I,1}P_{2I,3};D_{2I,3},D_{2I,5}; F_{2I,5}, F_{2I,7}\, ;\nonumber\\
&&G_{2I,7}, G_{2I,9};H_{2I,9},H_{2I,11}\, ,\quad
\mbox{for}\quad I= {1\over 2},\,  {3\over 2}\, .
\label{class_scheme}
\end{eqnarray}
Note that the states with the highest spins $J^\pi = (k+{1\over 2})^\pi $ 
from the Lorentz multiplets under consideration and the corresponding 
isolated $\lbrace J^\pi,0\rbrace \oplus \lbrace 0, J^\pi \rbrace $ states 
may belong to different Fock spaces due to different origins
for their parities. Consider, as an illustration, a spin--3/2$^-$ resonance  
to emerge as the highest spin state of the Lorentz multiplet
$\lbrace 1/2,1/2\rbrace \otimes \lbrack\lbrace 1/2,0\rbrace\oplus
\lbrace 0,1/2\rbrace \rbrack $.  
The negative parity of this resonance signals an internal  $P$ wave state.
Such a natural parity state resides within a Fock space built upon a scalar 
vacuum. On the contrary, the negative parity of an isolated
$\lbrace 3/2^-,0\rbrace \oplus \lbrace 0, 3/2^-\rbrace $
state arises necessarily from the coupling of the fermion degree
of freedom, when considered as fundamental, to an internal $0^-$ state. 
Such an unnatural parity state necessarily resides within a Fock space built 
upon a pseudoscalar vacuum. This aspect of mapping isolated higher--spin 
states onto multi--spinor representations of the Lorentz group is of crucial 
importance for selecting baryons as genuine parity partners and, therefore, 
for establishing the scale of chiral symmetry restoration (see below).

Considering resonances as approximate O(4) partial waves 
brings numerous advantages relative treating them
as O(3) partial waves. First of all, the number of
the  `missing resonances' is strongly reduced. 
Indeed, within that scheme the $\Delta $ 
spectrum below 2000 MeV appears complete as all observed
states fit into the $l'=0$, and $l'=1$ Lorentz multiplets in 
Eq.~(\ref{Wbrg_spinor}).
As the F$_{37}$ state from the $l'=1 $ Lorentz multiplet has to be 
paralleled in the nucleon sector by a (not reported) F$_{17}$ resonance 
with a mass around 1700 MeV, only that latter state has to be viewed as a 
`missing resonance' among the non-strange baryon excitations with masses 
below 2000 MeV.
In continuing by comparing the states from the third nucleon and 
$\Delta $ clusters with $l'$=2, four more `missing resonances' are predicted. 
These are the H$_{1, 11}$, P$_{31}$, P$_{33}$, and D$_{33}$ states with 
masses between 2200 and 2400 MeV. 
In summary, five new, still unobserved non-strange resonances have been 
predicted in \cite{Ki1}. Remarkably, the spacing of about
200 MeV among the relativistic multiplets with masses below 2000 MeV 
appears now much larger as compared to the maximal mass splitting of 
50-70 MeV between the corresponding multiplet members (see Fig. 1). 
The baryon classification scheme in Eq.~(\ref{class_scheme})
shows that the parities of the resonances from the first spin--clusters with 
$l'=0$  are always natural. This means that they can be considered to emerge 
within a Fock space built upon a vacuum of positive parity as spontaneously 
selected in the course of the realization of chiral symmetry in the hidden 
Nambu--Goldstone mode in the low energy regime.
\begin{figure}[htbp]
\centerline{\psfig{figure=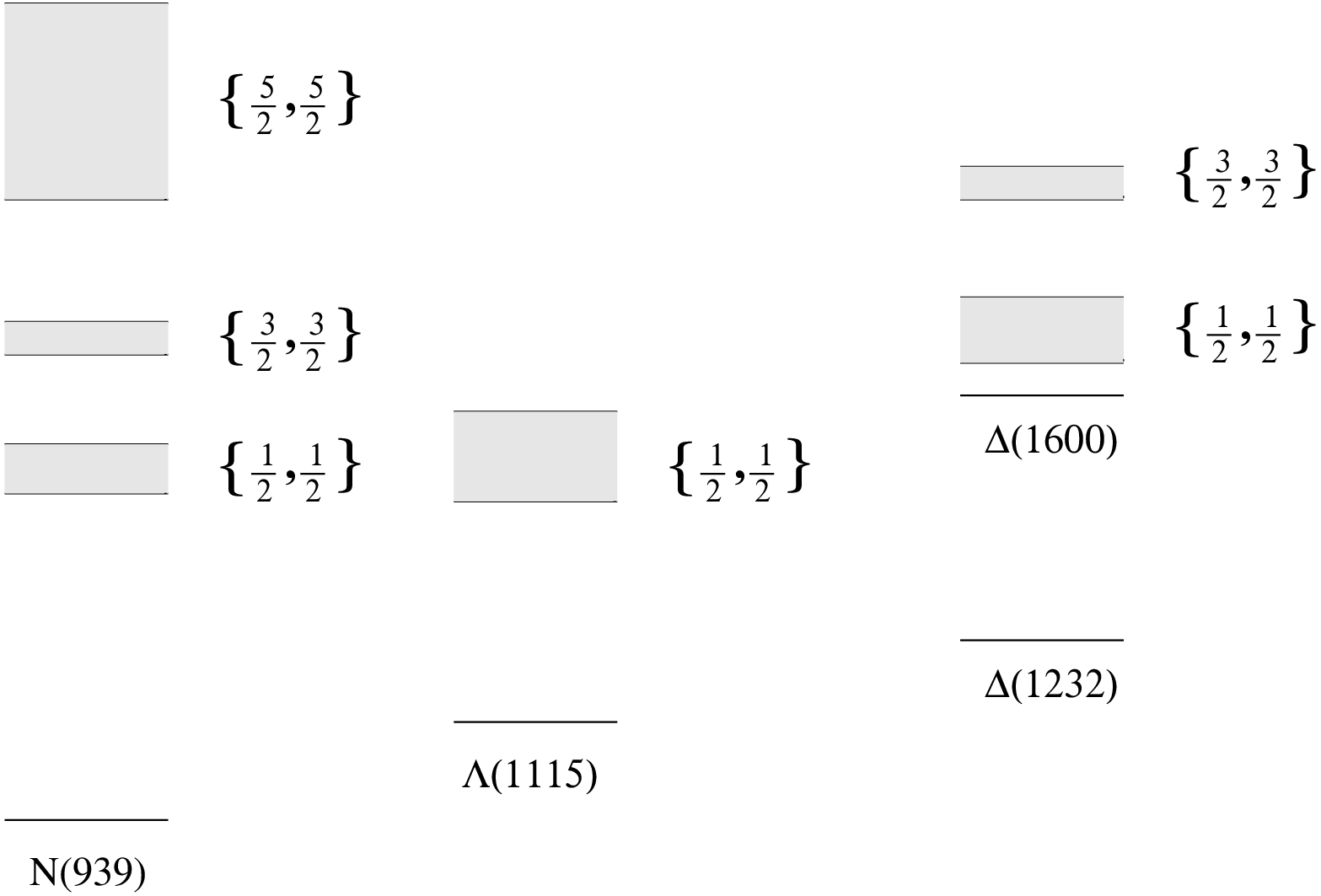,width=10cm}}
\vspace{0.1cm}
{\small Fig. 1\hspace{0.2cm} 
Baryon spectra in terms of the
$\lbrace j,j\rbrace \otimes 
\lbrack \lbrace {1\over 2},0\rbrace \oplus \lbrace 0, {1\over 2}
\rbrace \rbrack $ representations.
Only the $\lbrace j,j\rbrace $ 
assignments are given. Each dashed area contains
$4j+1 $ states. 
 }
\end{figure}
On the contrary, the resonances of the second and third
spin--clusters with $l'=1$, and $l'=2$, respectively, have always unnatural
parities and can be considered to reside within 
a Fock space built upon a vacuum of negative parity.
Both the $\psi_N\to \Psi_{\mu\mu_1\mu_2}$, and
$\psi_N\to \Psi_{\mu\mu_1...\mu_4}$ transitions are, therefore,
in contrast to the $\psi_N \to\Psi_\mu $ excitation, 
chiral phase transitions. A further advantage of the 
O(1,3)$_{LS}\otimes $SU(2)$_I$ classification scheme for baryons is that it 
opens a new possibility for the relativistic description
of the higher-spin resonances as members of covariant representations.
Indeed, the relativistic propagators 
of the spin--clusters can be directly read off from the representation 
theory of the Lorentz group as
\begin{equation}
S_{\mu\mu_1...\mu_{k-1}\,  ;\, \nu\nu_1 ...\nu_{k-1}}={
{ (\gamma\cdot\, p  +M ) 
\bigotimes_{n=1}^{n=k}  (g_{\mu_n\nu_n} - {1\over M^2}
p_{\mu_n}p_{\nu_n}) }\over
{2M(p^2-M^2)} }\, , 
\label{clstr_prop}
\end{equation}
with $M$ standing for the mass of the degenerate resonances.
For example, the ($S_{2I, 1}, D_{2I, 3}$ ) cluster is described 
in terms of the Lorentz vector with Dirac spinor components
$\Psi_\mu $ from Eq.~(\ref{class_scheme}) 
and its propagator is given by \cite{Ki1} 
\begin{equation}
S_{\mu\nu} = {{(\gamma\cdot\, p +M)(g_{\mu\nu}
-{1\over M^2} p_\mu p_\nu  )}
\over {2M(p^2-M^2)}}\, .
\label{cluster_prop}
\end{equation}
In noting that, say, the first S$_{11}$ and D$_{13}$ 
states appear separated by only 15 MeV, one sees that the relativistic 
contribution of these states to the amplitude of processes like meson 
photoproduction at threshold, can easily be calculated. Along the line 
of the  representation theory of the Lorentz group, the construction of
higher-cluster propagators and interactions with external fields
is straightforward. For example,
for the case of a ${\cal B}\to N +V$ process, where 
${\cal B}$ stands for a Lorentz covariant spin-cluster, 
while $V$ is a vector meson, a possible Lagrangian can be written as 
\begin{eqnarray}
{\cal L}_{{\cal B}VN} = 
\bar\Psi ^{\mu\mu_1...\mu_{k-1}}
({f_k\over m_\pi^{k-1}}\partial_{\mu_1}...\partial_{\mu_{k-1}} A_{\mu}
&+&  {{f_k'}\over m_\pi^{k} }
\partial_\mu ... \partial_{\mu_{k-1}} A\!\!\! / )\, 
\psi_N\, , \end{eqnarray}
where $k=2l'+1$, $A_\mu $ denotes the vector meson field, while
the coupling strengths  $f_k $ and $f_k' $ can be extracted from
data.

The Lorentz covariant spin-clusters of baryon resonances introduced above
are in fact nothing else but the special case of the Rarita--Schwinger 
representations with odd $k$'s, the essential difference being that 
all the lower-spin states are no longer 
redundant components that need be eliminated, but 
{\it physically observable resonances\/} related to the
composite character of the baryons. 
The Lorentz--Dirac index notation for the spin--clusters introduced
in Eq.~(\ref{class_scheme}) indicates that 
the first resonance-tower will predominantly couple to the 
pion--nucleon (or $\eta $--nucleon) system carrying each one Dirac
and one Lorentz index. The second and third spin--clusters
will prefer couplings to multipion--nucleon final states
(one Dirac- and several Lorentz indices) in agreement with the empirical 
observations. According to that,
the reason for the suppression of the S$_{11}$(1650)$\to $N+$\eta $
decay channel as compared to the S$_{11}$(1535)$\to $N+$\eta $ one,
can be a simple re--distribution of decay strength in favor of the
new opened S$_{11}$(1650)$\to $N+$\pi $+$\pi$ channel in accordance with
data.
\noindent
Now, the existence of the Lorentz covariant spin--grouping
of baryon resonances can be interpreted to emerge via
the coupling of a quark to approximately mass--degenerated Di-quarks of 
spins $l=0,..., 2l'+1 $ having all either natural or unnatural parities. 
Indeed, the covariant quark-Di-quark model based on solving the
Bethe-Salpeter equation reveals an internal O(4) symmetry \cite{Kus}
as visible from the rapid convergence of its solutions in the basis
of the orthogonal (Gegenbauer) polynomials of the
group O(4) corresponding to four dimensional partial waves.

Thus the major advantage of our new relativistic spectrum generating 
algebra for baryons is that it reconciles such seemingly 
contrary ideas of the baryon dynamics like the constituent quark
model on the one side, and the multi--spinor representations of
the Lorentz group corresponding to structureless (pointlike) particles, 
on the other side. 
Now the apparent analogy between the spectrum of the hydrogen atom
and the baryon spectra leads to the question whether 
the positions of the Lorentz covariant spin--clusters 
is determined by the inverse squared $1/n^2$ of the principle quantum number
and follow a sort of modified Balmer-series pattern. 
The answer to that question is positive \cite{Ki1}. Below quite a simple  
empirical recursive relation is suggested that describes with quite an 
amazing accuracy the reported mass averages of the resonances
from the Lorentz multiplets  only in terms of the cluster quantum numbers 
and the masses of the ground state baryons:
\begin{eqnarray}
M_n -M_{n-2}
&=&
\left( {1\over {(n-1)^2}}-{1\over {n^2}}\right)
 J_{n-2}^{max} \left(J_{n-2}^{max} +1 \right) M_{n-2} \, ,
\nonumber\\
n &=& 2l' +2 
\, , 
\qquad l'>0\, ,  \quad J_{n-2}^{max} = n-{1\over 2}. 
\label{Balmer_ser}
\end{eqnarray}
Here, M$_n$ denotes the mass of the respective Lorentz multiplet.
The position of the first excited nucleon spin--cluster with $l'=0$ is
given instead by (\ref{Balmer_ser}) rather by
$ M_{n=2}- M_{n=1} =  ({1\over 1^2} -{1\over 4})\,
(1^2 -{1\over 4})M_{n=1}$ with $M_{n=1}=M_N$. 
For the $\Delta $ and $\Lambda $ baryons, the mass 
scale $M_{n=1}$, has to be replaced by 
$ (M_N+M_\Delta )/2$, and $ (M_N+M_\Lambda )/2$, respectively
(compare Table 2 of Ref.~\cite{Ki2} for details). To conclude, 
baryon resonances group to O(4) rather than to O(3) partial waves,
and the baryon spectra are evidently generated by
the relativistic o(1,3)$_{LS}\otimes $su(2)$_I$ group
algebra rather than by the algebra of the non--relativistic
O(3)$_{L}\otimes $SU(6)$_{SF}$ group

Work supported by the Deutsche Forschungsgemeinschaft (SFB 201).


\begin{thebibliography}{99}
\bibitem{Kleinert} H.\ M.\ Kleinert,  Fortschritte der Physik,
                   {\bf 16}, 1 (1968).    
\bibitem{Toller} M.\ Toller, Nuovo Cim.\ {\bf 54A}, 295 (1968). 
\bibitem{J00J} S.\ Weinberg,  Phys.\ Rev.\  {\bf 133}, B1318 (1964). 
\bibitem{Hoehler} G.\ H\"ohler, In: {\it Physics with GeV- Particle Beams\/}
                   eds. H.\ Machner and K.\ Sistemich (World Scientific, Singapore)
                   1995, p. 198.
\bibitem{Ki1} M.\ Kirchbach, Mod.\ Phys.\ Lett.\ {\bf A12}, 2373 
                    (1997); ibid. 3177. 
\bibitem{Ki2} M.\ Kirchbach, Mod.\ Phys.\ Lett.\ {\bf A13}, 823 (1998).
\bibitem{Kus} K.\ Kusaka {\it et al.,\/} Phys.\ Rev.\  {\bf D55}, 5299 (1997).
\bibitem{Part} Particle Data Group, R.\ M.\ Barret {\it et al.,}
                Phys. Rev.\ {\bf D54}, 1 (1996).
\end{thebibliography}
\end{document}